# Fractional Authorship in Nuclear Physics

B. Pritychenko

National Nuclear Data Center, Brookhaven National Laboratory, Upton, NY 11973-5000, U.S.A.
E-mail: pritychenko@bnl.gov, Phone: (631) 344-5091, Fax: (631) 344-2806

**Abstract** Large, multi-institutional groups or collaborations of scientists are engaged in nuclear physics research projects, and the number of research facilities is dwindling. These collaborations have their own authorship rules, and they produce a large number of highly-cited papers. Multiple authorship of nuclear physics publications creates a problem with the assessment of an individual author's productivity relative to his/her colleagues and renders ineffective a performance metrics solely based on annual publication and citation counts. Many institutions are increasingly relying on the total number of first-author papers; however, this approach becomes counterproductive for large research collaborations with an alphabetical order of authors. A concept of fractional authorship (the claiming of credit for authorship by more than one individual) helps to clarify this issue by providing a more complete picture of research activities. In the present work, nuclear physics fractional and total authorships have been investigated using nuclear data mining techniques. Historic total and fractional authorship averages have been extracted from the Nuclear Science References (NSR) database, and the current range of fractional contributions has been deduced. The results of this study and their implications are discussed and conclusions presented.

# 1 Introduction

Nuclear and particle physics authorship is an exciting frontier of scientometrics research. A recent joint publication of ATLAS and CMS collaborations has set a new record for total number of authors per single research paper at 5,154 (Aad *et al.* 2015, Castelvecchi 2015). Such an extremely large list of authors reflects the large scale of modern scientific research and has become a new normal in high energy and nuclear physics.

Historically, scientific authorship has been about individual achievements, or substantial contributions to a joint project. A simple count of publications and corresponding citations has been used as a basic metrics for performance estimates of individual scientists. However, in recent years, the whole paradigm of nuclear research has changed, and large groups dominate the landscape (Pritychenko 2015). These groups run sophisticated experiments and analyze large volumes of data. Large-scale, groundbreaking research requires contributions from many people with different sets of skills who often work on small parts of large projects. Such a mode of research is well described by a popular business saying, `There Is No "I" In Team'. Simultaneously, there are many examples when routinely-repeated measurements are authored by large groups of researchers because of collaborative authorship policies.

The increase in authorship of nuclear physics publications creates a problem with the assessment of an individual author's productivity. A large number of multiple-author papers in high-impact journals obscure the picture, and such scientists dominate over people who work in small groups or alone. This makes it extremely difficult to judge an individual scientist's performance using the total number of publications or citations (Mohamed Gad-el-Hak 2004). Therefore, many scientific institutions worldwide are increasingly relying on the total number of first-author papers. Tenure track promotions in many organizations depend on first-author or single-author publications; however, this approach creates an unusual problem for large research collaborations where order of authors is often alphabetical (Academic authorship 2015). Recently, the promotion of a high-energy professor at a major American university was delayed because of the lack of first-author publications. It took additional paperwork and other efforts to convince tenure committee members that the scientist deserved the promotion. A concept of fractional authorship helps to resolve this issue by providing a more complete picture of research activities (Plume, Weijen 2014).

To investigate the above-mentioned dilemma, I will analyze the authorship of nuclear physics articles using the Nuclear Science References database (Pritychenko, Betak, Kellet, et al. 2011) maintained by the National Nuclear Data Center (NNDC) (Pritychenko, Herman 2012), calculate total and fractional average values, evaluate the ranges of fractional contributions, and deduce trends for the last century. These historic trends constitute a basis for individual performance metrics. Data analysis methods and corresponding results are presented in the following sections.

# 2 Nuclear Bibliography Data Mining

Nuclear bibliography materials represent the foundation for nuclear data compilations and evaluations. U.S. Nuclear Data Program and International Atomic Energy Agency data scientists systematically have collected these materials and compiled the database records over the past 65 years. The NSR database (http://www.nndc.bnl.gov/nsr) contains metadata for more than 219,000 publications, and 94,000 individual authors (Pritychenko, Betak, Kellet, et al. 2011). The references database provides almost complete coverage of nuclear physics publications and is relatively clean of high-energy physics papers, where authorship rules are very different (Academic authorship 2015). It is a prime source of nuclear bibliography worldwide and an excellent resource for investigation of authorship from 1896 to the present.

The database contents, as of September 2015, have been examined with nuclear data mining techniques. A custom-written Java code with embedded SQL queries has been employed to examine the database, process the metadata, and deduce historic averages, ranges, and trends. The results of this study are presented below. This database does not contain the complete publication records for individual scientists; it is rather the most complete highly-specialized bibliography of nuclear physics results. The NSR is missing many nuclear instrumentation and related topic articles due to database scope changes and technological reasons. The database absolute values may differ from other studies, and they rather reflect the lower limits for certain individual authors, while research evolution trends over the last 100 years are very reliable.

## 3 Historic Trends in Nuclear Physics Authorship

Historic evolution of the nuclear bibliography for the entire and experimental contents is shown in the upper and lower panels of Fig. 1, respectively. The upper panel of the Figure clearly shows a World War II dip in nuclear physics publications, with an absolute minimum in 1944, due to atomic programs in the United States, the United Kingdom, Germany and the Soviet Union (Manhattan project 2015). While the lower panel strongly depends on NSR keywording data consistency to separate experimental publications, its step-function-like structures reflect the start of experimental nuclear physics campaigns in the 1930s and the different style of NSR keyword compilations by Oak Ridge Lab in the 1960s. Both panels provide complementary information on the size of statistical samples and nuclear physics trends.

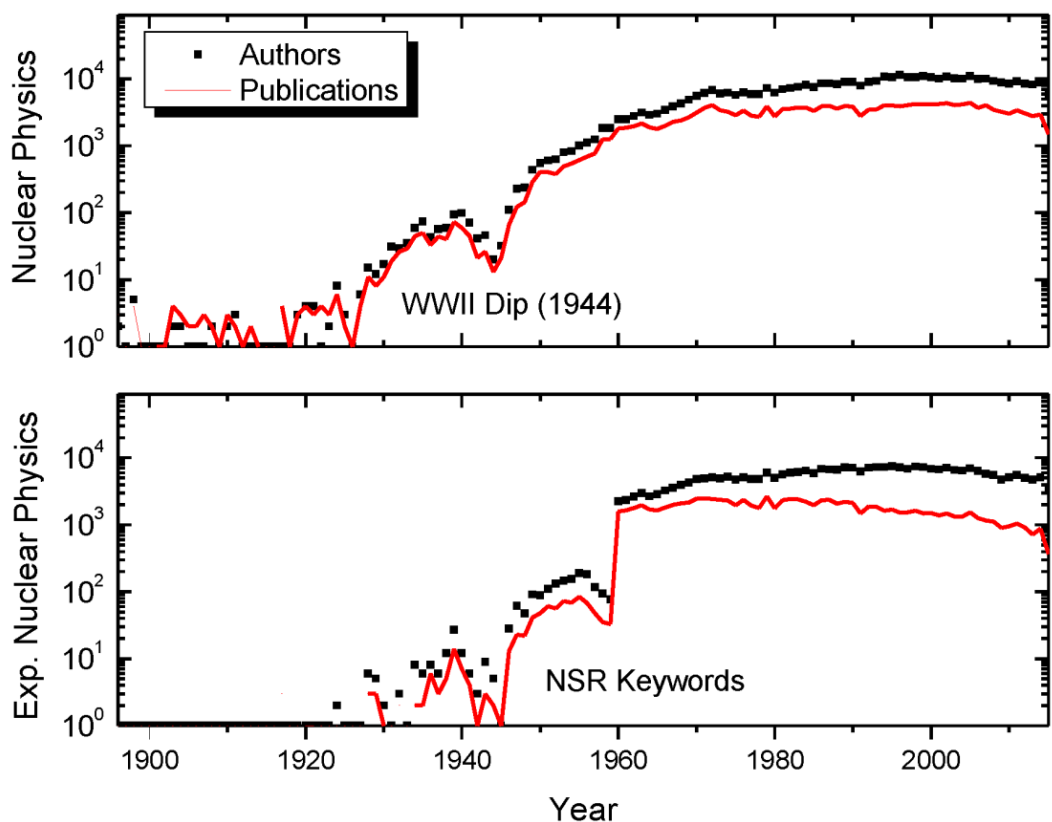

**Fig.1.** Annual bibliography trends for the entire and experimental nuclear physics contents. Data were taken from the Nuclear Science References database.

Previous sections indicate that a performance metrics that is solely based on total number of publications and citations per year is not very useful for large experimental groups, and additional performance criteria are needed. To extend the scope of these findings, I will split the data sample into smaller, single-author bins and analyze fractional authorship in the last 100 years. Fractional authorship quantifies an individual author's contributions to published papers, and it can be used to estimate the number of papers written by a particular scientist.

Following the pattern described by Plume and Weijen, I would analyze the evolution of individual contributions using NSR bibliographic data. Fig. 2 shows fractional authorship averages from the NSR database. The upper panel of the Figure clearly indicates contrasting trends between total and fractional numbers of publication averages. Initially, single-author publications cause the curves to overlap, but later, total averages are increasing while fractional authorship is in a decline. Furthermore, relatively flat fractional authorship curves clearly show that individual productivity stayed roughly the same in the last 50 years. This result agrees with the recent analysis of Elsevier publications (Plume, Weijen 2014). The contrasting trends are more dramatic for experimental nuclear physics papers as shown in the lower panel of Fig. 2. The insufficient statistics affect calculated averages at the beginning and end of the time scale, and the NNDC is allocating additional resources for recovery of missing nuclear physics publications.

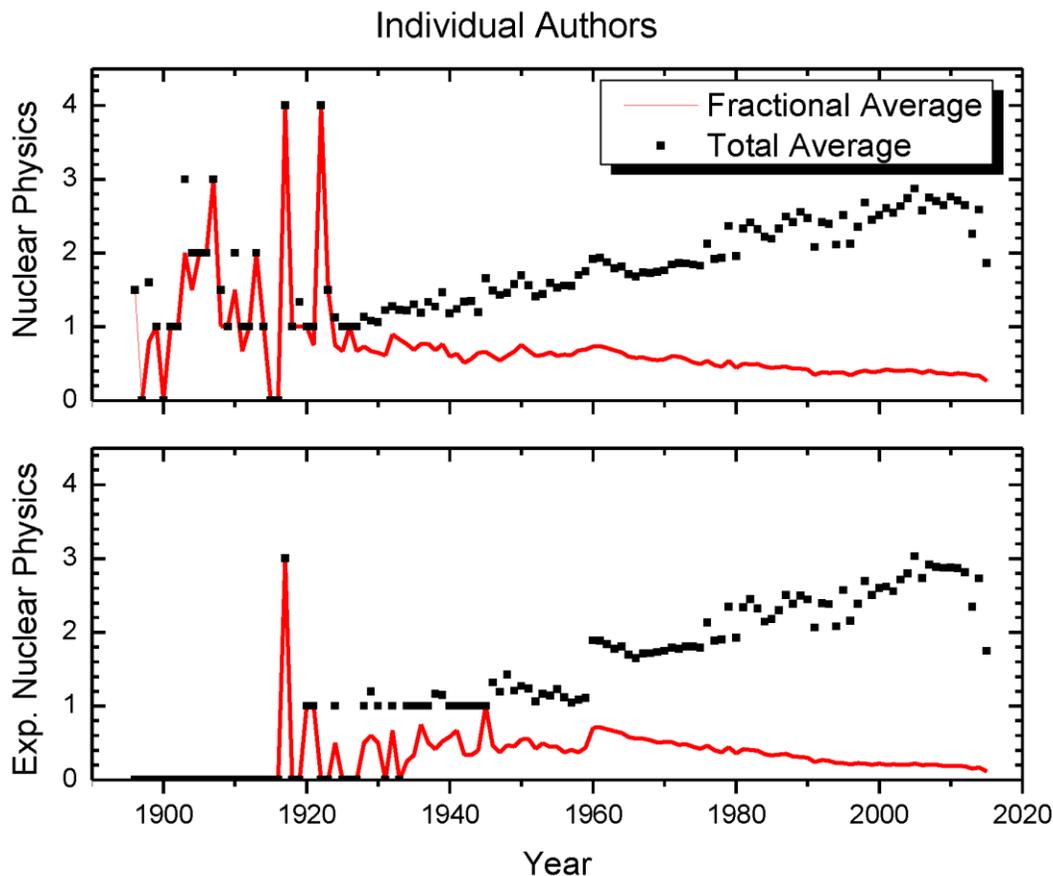

**Fig.2.** The individual author total number of publications and fractional authorship averages in the entire and experimental nuclear physics. Data were taken from the Nuclear Science References database.

To estimate the ranges of nuclear physics authorship, the current analysis can be applied to the most prolific NSR author that has been credited with 1,038 publications. The last number can be compared to 603 Google Scholar's retrievals for a profile of 1962 Nobel Prize winner Lev Landau (1908-1968), who was famous for his universal approach and theoretical physics textbooks translated into many languages (Google Scholar 2015, Kojevnikov 2015). The Google Scholar count is inflated by the 262 posthumously-published books, while the total number of independent works is only 123 (Complete list of L D Landau's works 1998). Currently, we have at least 12 NSR authors with a publication count higher than the Google Scholar count for Landau; authorship rules definitely have evolved in the last 40-50 years, and NSR authors could accumulate large numbers of publications because of lenient nuclear physics authorship rules when even minor contributions or participation in experimental campaigns could justify credit. These lenient authorship rules can be compared with an American trophy culture when all the sports players get awards (Banks 2015). Scientific authorship should not be merely a participation award; it should be a reflection of substantial contributions. It is supposed to represent some degree of individualism in successfully-accomplished projects because `There is no "I" in team but there is in win' (Jordan 2009).

A complementary calculation and analysis of fractional contributions for the NSR's most prolific author are presented below. The fractional contributions are estimated by using two methods: a simple arithmetic and first-author fractions. The arithmetic fraction is based on an equal distribution of authorship credit among coauthors, while in nuclear physics, it is reasonable to assume that the first author usually contributes ~50%, and the rest is split between the other coauthors. Single- or double-author papers are credited identically in both cases. These rules reduce 1,038 publications over 37 years to 78.54 and 54.48 publications for arithmetic and first-author fractions, respectively. The simple arithmetic fraction count is more favorable for the most prolific NSR author because he was a leading author in 25 out of 1,038 cases, including four single-author papers. Further examination of Fig. 3 indicates that only in a few cases does fractional authorship exceed four publications a year, while the total number of publications always has been in double digits. The current example clearly demonstrates the value of fractional authorship for realistic estimates of an individual scientist's productivity; a complementary finding shows overall productivity has not been affected by computer technologies over the years.

The current data clearly sets an upper boundary for fractional nuclear physics contributions at 4-5 publications/year; the lower boundary is defined by the ATLAS and CMS collaborations paper (Aad *et al*. 2015, Castelvecchi 2015). Individual averages for the entire and experimental nuclear physics are ~0.4 and ~0.2 publications/year, respectively, as shown in Fig. 2. These nuclear physics values are consistent with the Elsevier fractional average value of 0.56 (Plume, Weijen 2014).

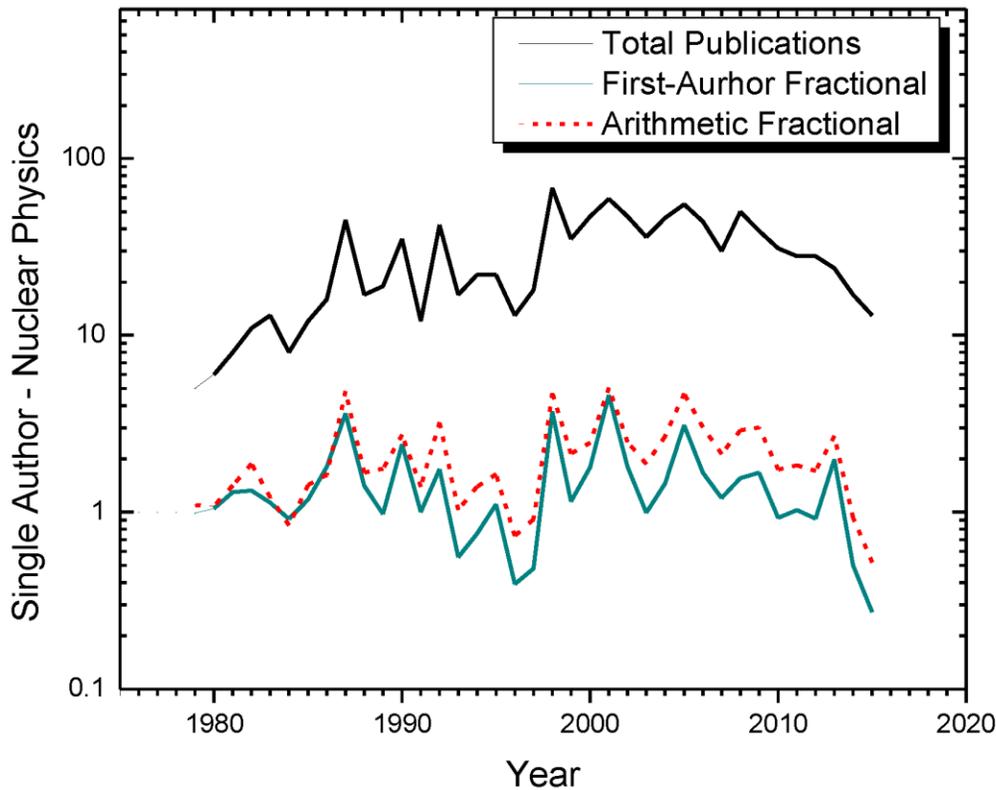

**Fig.3.** Single-author annual trends for the total number of publications and first-author and arithmetic fractional authorship averages in nuclear physics. Data for the most prolific author were taken from the Nuclear Science References database.

Further analysis of the 93,867 nuclear physics authors shows only 43,037 individuals who have published first-author papers, including 35,338 non-alphabetical author lists. These results highlight two distinct groups of authors: project leaders (risk takers) and followers. Such a big discrepancy between two groups cannot be explained by an alphabetical order of authors in former Soviet Union countries because their total contribution is less than 15% (Pritychenko 2015). These numbers rather indicate excessive author lists, and the fact that many graduate students are leaving the science. Nuclear physics laboratories are training people for many other occupations.

## 4 Interesting Nuclear Science Patterns

The present data analysis also shows many interesting patterns in nuclear physics research. Authorship rules in large groups differ from the traditional and result in uncommon approaches (Goodman 2012). Larger than normal author lists are often observed in high-impact journal publications such as Physical Review Letters. These groups have developed new, creative ways of collaborating on joint publications. Fig. 4 shows a screenshot of a LaTex template of a D0 collaboration where the list of authors is maintained in a separate author_list.tex file (Template for PRL/PRD Papers - Fermilab 2015). It helps the article writers concentrate on the content instead of constant updates of the contributing authors and organizations. This innovation clearly simplifies the publication process for large collaborations with thousands of coauthors.

```
\begin{document}

% The following information is for internal review, please remove them for submission
\widetext
\leftline{Version xx as of \today}
\leftline{Primary authors: Joe E. Physics}
\leftline{To be submitted to (PRL, PRD-RC, PRD, PLB; choose one.)}
\leftline{Comment to {\tt d0-run2eb-nnn@fnal.gov} by xxx, yyy}
\centerline{\em D\O\ INTERNAL DOCUMENT -- NOT FOR PUBLIC DISTRIBUTION}

% the following line is for submission, including submission to the arXiv!!
%\hspace{5.2in} \mbox{Fermilab-Pub-04/xxx-E}

\title{Template for PRL/PRD Papers}
\input author_list.tex          % D0 authors (remove the first 3 lines
                                 % of this file prior to submission, they
                                 % contain a time stamp for the authorlist)
                                 % (includes institutions and visitors)
\date{\today}

\begin{abstract}
An article usually includes an abstract, a concise summary of the work
covered at length in the main body of the article. It is used for
```

**Fig.4.** Physical Review Letters & Physical Review D LaTex template for high energy physics papers (Template for PRL/PRD Papers - Fermilab 2015).

Unfortunately, a large number of authors does not always guarantee the quality of published materials. A recent Physical Review Letters publication (Lorusso, Nishimura, Xu et al. 2015) passed through the editorial screening and referees, and was successfully published in this very prestigious journal. A seven-page long and 55 author-strong article has one major and four minor typos with measured decay half-lives, inconsistency between table data and plots, and ignores the possible contribution of isomeric state decays. These problems could have been avoided if all authors would have read the paper carefully before submission to the journal.

Similar issues have been present in other papers. A debacle with the discovery of elements of Z=116 and 118 (Ninov, Gregorich, Loveland, et al. 1999) sheds more light on issues with many-author nuclear physics publications. A single person analyzed the groundbreaking experiment, and none of the14 coauthors double-checked the findings. Subsequent failures to reproduce the results by other researchers have triggered the re-analysis of the original data, and scientific fraud has been discovered (Ninov, Gregorich, Loveland, et al. 2002, Davidson 2002). This data manufacturing by a lead author would not have happened if the group members had considered not just authorship rights but also its responsibilities more seriously. Large number of authors could have been a blessing instead of curse for these papers if physicists had worked together on all stages of the research project.

Obviously, these cases are exceptional and do not correctly represent nuclear physics research operations. In fact, the Ninov's fiasco has forced scientists to modify the rules for new element discovery by requiring complete α-decay chain observation. However, occasional unwillingness to read their own papers and assume ownership of the data by several physicists (Goodman 2012) provides an additional rationale for introduction of fractional authorship credit in nuclear physics.

## 5 Conclusions

Recent developments in experimental nuclear physics authorship defy the traditional rules of individual performance assessment for nuclear scientists. They render obsolete performance metrics solely based on annual publication and citation counts. A complementary criterion of fractional authorship helps to remedy this issue and improves the metrics. This additional metrics parameter reflects a number of written papers by individual authors and solves discrepancies between small and large author list publications. It helps judge a scientist's performance according to his/her individual contributions.

Fractional authorship in nuclear physics has been investigated using the nuclear data mining of relational database contents. Historic averages for total and fractional authorship have been examined and the current range for fractional contributions has been deduced. Further analysis of averages indicates an anti-correlation effect between an increase in total number of publications and decrease in fractional authorship, as well as relatively flat individual productivity in recent years. Additional research is necessary to investigate these findings in other areas of science.

## 6 Acknowledgments

The author is grateful to U.S. Nuclear Data Program members and Dr. V. Unferth (Viterbo University) for productive discussions and careful reading of the manuscript and useful suggestions, respectively. This work was funded by the Office of Nuclear Physics, Office of Science of the U.S. Department of Energy, under Contract No. DE-AC02-98CH10886 with Brookhaven Science Associates, LLC.